\documentclass[twocolumn,aps,prd,showpacs,nofootinbib,superscriptaddress,floatfix,showkeys]{revtex4-2}

\usepackage{amsfonts}
\usepackage{amsmath}
\usepackage{amssymb}
\usepackage{bm}
\usepackage{dcolumn}
\usepackage{graphicx}   
\usepackage[latin1]{inputenc}
\usepackage{latexsym}
\usepackage{rotating}
\usepackage{hyperref}
\usepackage{graphicx}
\usepackage{color}

\newcommand\be{\begin{equation}}
\newcommand\ba{\begin{eqnarray}}
\newcommand\ee{\end{equation}}
\newcommand\ea{\end{eqnarray}}

\newcommand{\gcoup}{g_{\phi \gamma}}
\newcommand{\bo}{\mathbf}

\begin{document}

\title{Secondary Production of Photons from ALP Dark Matter interacting with a Cosmological Magnetic Field}

\author{Abdias Aires}
\affiliation{Department of Physics, McGill University, Montr\'{e}al,
  QC, H3A 2T8, Canada}
\affiliation{Instituto de Fisica Teorica, UNESP-Universidade Estadual Paulista
R.  Dr.  Bento T. Ferraz 271, Bl. II, Sao Paulo 01140-070, SP, Brazil }

\author{Robert Brandenberger}
\email{rhb@physics.mcgill.ca}
\affiliation{Department of Physics, McGill University, Montr\'{e}al,
  QC, H3A 2T8, Canada}
\affiliation{Trottier Space Institute, Department of Physics, McGill
University, Montr\'{e}al, QC, H3A 2T8, Canada}

\author{Ashu Kushwaha}
\email{kushwaha.a.celb@m.isct.ac.jp}
\affiliation{Department of Physics, Institute of Science Tokyo,
2-12-1 Ookayama, Meguro-ku, Tokyo 152-8551, Japan}


\begin{abstract}
	
 Under the assumption that dark matter is a coherently oscillating pseudoscalar field coupled to electromagnetism by the usual Chern-Simons term, we study the production of secondary photons from dark matter fluctuations coupled to a pre-existing magnetic field, taking into account the spectral distribution of the magnetic field. Specifically, we apply the formalism to the case of a large-scale magnetic field generated previously via a parametric resonance instability due to the same Chern-Simons coupling.  However, our analysis is applicable to any spectrum of cosmological scale magnetic field fluctuations present at the time of recombination. We show that obtaining a sufficiently large flux of photons in the Lyman-Werner frequency range is consistent with constraints from CMB and X-ray observations.

\end{abstract}

\maketitle

\section{Introduction} 
\label{sec:intro}

There has recently been a growing interest in coherently oscillating ultra-light axion-like particles (ALPs) as the dark matter (see \cite{original} for the original articles on axions as dark matter, and \cite{ALPrev} for reviews on ALP dark matter).  Such particles can couple to electromagnetism via a Chern-Simons term.  Effects of this coupling have been studied in many works (e.g. \cite{earlier}). Recently, it was pointed out \cite{BFJ} (see also \cite{earlier2}) that, in the absence of plasma effects, this coupling can induce a parametric resonance instability \footnote{The same coupling has been used extensively to study magnetic field generation during inflation \cite{inflation-B}, and in \cite{Namba} a possible parametric resonance instability was pointed out.} which can lead to the generation of cosmological magnetic fields immediately after the time of recombination, when plasma effects become suppressed \footnote{It remains an open question to study the role of the residual small ionization of space in suppressing the resonance \cite{residual}.  Following the justification given in \cite{Nirmalya}, we will assume that the resonance process is robust.}.  The resonance can either be of tachyonic type \cite{BFJ} or else a narrow resonance instability \cite{Nirmalya}, depending on the values of the parameters in the action: the ALP mass and the coupling constant in the Chern-Simons term.  Note that the resonance produces long wavelength modes of the electromagnetic field - it is an infrared instability \footnote{The mechanism of \cite{BFJ} has been generalized to scalar dark matter \cite{Kamali} and vector dark matter \cite{Tatsuya}.}.

In \cite{Jiao} the possible role of this magnetic field in opening up the direct collapse black hole formation channel for the formation of super-massive black holes was first mentioned.  For this, a high flux of photons in the Lyman-Werner (LW) range is required (see e.g. \cite{Direct-Collapse} for an original work on the direct collapse black hole formation scenario, and \cite{SMBH}). However,  there are obstacles to the cascade and thermalization arguments which were used in \cite{Jiao} to obtain the required LW flux.  In a recent paper \cite{Ashu} it was pointed out that there is a natural way to generate secondary photons from a pre-existing magnetic (B) field, namely by coupling the B-field to the dark matter density fluctuations (which are realized as fluctuations of the ultralight pseudoscalar field $\phi$). It was shown that a magnetic field on cosmological scales with a strength comparable to the observational lower bounds (see e.g. \cite{Durrer} for a review article on magnetic fields, and \cite{Vovk} for studies of the lower bound which comes from cosmic ray observations) is able to produce a sufficiently high flux of LW photons to prevent molecular hydrogen formation and hence to allow for direct collapse super-massive black hole formation.

In \cite{Ashu} the magnetic field was taken to be homogeneous. This is a good approximation if we are interested in the production of secondary photons with microphysical wavelength such as the LW photons.  However, the mechanism also leads to the generation of secondary photons with much larger wavelengths, e.g. wavelengths which are relevant for cosmic microwave background (CMB) spectral distortions. For such wavelengths the approximation of a homogeneous magnetic field breaks down. 

In this letter we compute the spectrum of secondary photons in the case of an inhomogeneous magnetic field defined by its power spectrum $P_B(k)$, where $k$ indicates comoving wave number. Specifically, we have in mind a spectrum which is peaked at some infrared scale $k_c$, but which is extended in $k$.  We compute the induced spectrum in the X-ray region and on scales relevant for CMB spectral distortions, and derive constraints on the parameter values for which the results are consistent with present limits assuming that the mechanism is able to yield a flux of Lyman-Werner photons which is sufficiently large to allow for direct collapse black hole formation.

Note that our analysis does not depend on the magnetic field having been produced by the parametric resonance mechanism of \cite{BFJ} and \cite{Nirmalya}.  We can consider any magnetic field present close to the time of recombination on cosmological scales. The spectrum of the magnetic field will, obviously, depend on the specific production mechanism.  With improved observations,  one will be able to use the observed magnetic field on cosmological scales as input into our calculations.

\section{Model}

We consider a standard Lagrangian which describes the coupling of an ultralight pseudoscalar dark matter field $\phi$ to electromagnetism
\be \label{Lag}
{\cal{L}} \, = \, \frac{1}{2} \partial_{\mu} \phi \partial^{\mu} \phi - \frac{1}{2} m^2 \phi^2 - \frac{1}{4} F^2 - g_{\phi, \gamma} \phi F \wedge F \, ,
\ee
where $F$ is the field strength tensor of electromagnetism, and $g_{\phi \gamma}$ is a coupling constant with dimensions of inverse mass.  If $\phi$ is to constitute the dark matter of the Universe, $\phi$ should start out oscillating coherently over cosmological scales. These coherent oscillations can be set up via a  misalignment mechanism, e.g. the one recently suggested in \cite{misalign}. 

As in \cite{BFJ} and subsequent work, we will parametrize the mass and the coupling constant in terms of dimensionless constants $m_{20}$ and ${\tilde g_{\phi \gamma}}$. The dimensionless mass is defined by
\be
m \, \equiv \, m_{20} 10^{-20} {\rm eV} \, ,
\ee
where, assuming that $\phi$ constitutes all of the dark matter of the Universe, observational bounds indicate that $m_{20} > 10$ \cite{Dalal}.  The dimensionless coupling constant is set by
\be
g_{\phi \gamma} \, \equiv \, {\tilde g_{\phi \gamma}} 10^{-10} {\rm GeV}^{-1} \, ,
\ee
where observational bounds indicate that ${\tilde g_{\phi \gamma}}$ cannot be much larger than unity (see \cite{ALPbound} for a discussion of the bounds). 

We consider initial conditions in which the electromagnetic gauge field $A_{\mu}$ starts out in its quantum mechanical ground state, and in which $\phi$ is coherently oscillating. As discussed in \cite{BFJ}, the coherent oscillations will, in the absence of plasma effects, lead to an infrared instability which can generate magnetic fields on cosmological scales. The resonance is either of tachyonic type \cite{BFJ} or of narrow resonance type \cite{Nirmalya}.  Before recombination,  plasma effects quench the instability, and thus the resonance sets in right after the time of recombination (modulo possible effects of the residual ionization \cite{residual}).
 
 Once the resonance is shut off and a cosmological magnetic field $B_i$ is generated, then, as discussed in \cite{Ashu}, a spectrum of secondary photons can be generated from the fluctuations in the dark matter field $\phi$. In the absence of a background electric field, the equation of motion for the secondary gauge fields (in Coulomb gauge) is
 \be \label{basicEoM}
 A_i^{\prime \prime} - \nabla^2 A_i \, = \, S_i \equiv - g_{\phi \gamma} \phi^{\prime} B_i \, ,
 \ee
 where a prime denotes the derivative with respect to conformal time and $\nabla$ is the d'Alembertian in comoving spatial coordinates.  In \cite{Ashu}, this equation was solved mode by mode in Fourier space assuming that $B_i$ is homogeneous. This is a reasonable approximation if we are studying the production of high frequency modes, but not if we are interested in lower frequency photons. If $B_i$ is homogeneous, then the Fourier mode $A_i(k)$ (k being the comoving wave number) is seeded directly by the k'th fluctuation mode $\delta \phi_k$ of the dark matter field. 
 
 If the magnetic field is not homogeneous, and instead is described by a Fourier mode distribution $B(k)$, then the source term $S_i(k)$ in (\ref{basicEoM}) is a convolution integral over all Fourier modes of $B(k')$ and $\delta \phi(k - k')$. This is the situation we consider here.
 
 We assume that the fluctuations in the dark matter field have a scale-invariant spectrum whose amplitude can be normalized to the Planck CMB anisotropy observations. As shown explicitly in \cite{Ashu}, the Fourier space density dark matter field fluctuations \footnote{Making use of the Fourier transform conventions discussed later in terms of which $\delta \phi(k)$ has canonical normalization.} have an amplitude
 \be \label{phik}
 \delta \phi(k) \, = \, \frac{1}{2} \Phi k^{-3/2} {\cal A}^{1/2} \, ,
 \ee
and oscillate with frequency $k$.  Here, ${\cal A}$ is the amplitude of the dimensionless power spectrum of the density fluctuations, and $\Phi$ is the amplitude of the $\phi$ oscillations. 

If $\phi$ constitutes the entirety of the dark matter at the time of recombination, then
\be
m^2 \phi^2 \, \sim \, T_{rec}^4 \, .
\ee
We shall use this relation in our later estimates.

\section{Flux of Secondary Photons}

The starting point is the equation (\ref{basicEoM}) for the secondary photons.  We use the following convention for the Fourier transform of a position space quantity $f(x)$:
\be
f({\bf x}) \, = \, V^{1/2} \int d^3k e^{i{\bf k} {\bf x}} f({\bf k}) \, ,
\ee
where $V$ is a cutoff volume (which will cancel out in all observables).  With this convention, the Fourier transform of a scalar field has canonical normalization.

If we Fourier transform (\ref{basicEoM}), we obtain the equation
\ba
A_i^{\prime \prime}({\bf k}) + k^2 A_i({\bf k}) \, &=& \, S_i(k) \\
& = & - g_{\phi \gamma} V^{1/2} \int d^3 {\bf {\tilde k}}
\delta \phi^{\prime}({\bf k} - {\bf {\tilde k}}) B_i({\bf {\tilde k}}) \, , \nonumber
\ea
where $k$ is the magnitude of the ${\bf k}$ vector. This equation can be solved using the Green's function method
\be \label{Aeq}
A_i(k, \eta) \, = \, \int_{\eta_{\rm rec}}^{\eta} d\eta^{\prime} G(\eta, \eta^{\prime}) S_i(k, \eta^{\prime}) \, ,
\ee
where $G(\eta, \eta^{\prime})$ is the Green's function of the homogeneous equation:
\be
G(\eta, \eta^{\prime}) \, = \, \frac{1}{k} {\rm sin}(k(\eta - \eta^{\prime})) \, .
\ee

\subsection{Magnetic field spectrum with a narrow peak }

We now assume that the spectrum of $B({\bf k})$ is isotropic in ${\bf k}$ but concentrated at a particular value of $k = k_c$:
\be \label{delta-spectrum}
B_i({\bf k}) \, = {\tilde B}_i(k_c) \delta (k - k_c) \, .
\ee
This assumption is natural if the magnetic field is produced via a narrow parametric resonance instability such as discussed in \cite{Nirmalya}.  If the magnetic field is produced via a tachyonic instability such as studied in \cite{BFJ}, the resulting spectrum is highly peaked at a particular value $k_c$, and thus the assumption (\ref{delta-spectrum}) is also a good approximation.

The first step in the computation is to express ${\tilde B}_i(k_c)$ in terms of the value $B$ of the magnetic field in position space. We find
\be \label{bvalue}
{\tilde B}(k_c) \, \simeq \, V^{-1/2} k_c^{-2} B \, .
\ee
Making use of (\ref{phik}) and (\ref{bvalue}), Eq. (\ref{Aeq}) then yields
\ba \label{result1}
A_i(k, \eta) \, &=& - g_{\phi \gamma} \frac{1}{2k} \Phi {\cal{A}}^{1/2} B \int_{\eta_{rec}}^{\eta}  {\rm sin}(k(\eta - \eta^{\prime})) d\eta^{\prime} \\
&\times& \int d^3 {\tilde{k}} |{\bf{k}} - {\bf{\tilde{k}}}|^{-1/2} {\rm cos}(|{\bf{k}} - {\bf{\tilde{k}}}| \eta^{\prime}) |{\bf{\tilde{k}}}|^{-2} \delta (|{\tilde{k}}| - k_c) \, . \nonumber 
\ea
Here, we have neglected the time dependence of the amplitude of oscillation of $\delta \phi$. This is justified provided that $m$ is smaller than $|{\bf{k}} - {\bf{\tilde{k}}}|$ for the range of ${\bf{\tilde{k}}}$ which enter the integral. 

Based on (\ref{result1}), the order of magnitude of $A_i(k)$ can be estimated to be
\be \label{large-limit}
A_i(k) \, \sim \, g_{\phi \gamma} \Phi {\cal{A}}^{1/2} B \frac{1}{k^{5/2}}
\ee
for $k \gg k_c$, while for $k \ll k_c$ we have
\be \label{small-limit}
A_i(k) \, \sim \, g_{\phi \gamma} \Phi {\cal{A}}^{1/2} B \frac{1}{k^2} \frac{1}{k_c^{1/2}} \, .
\ee
If the primordial magnetic field is generated via the mechanism of \cite{BFJ} or \cite{Nirmalya}, then $k_c \sim m$ for narrow resonance and $k_c \sim m \Phi g_{\phi \gamma}$ for tachyonic resonance.  The corresponding length scales are very small compared to the scale of CMB anisotropies, but very large compared the $X-ray$ wavelengths.

\subsection{A more general magnetic field spectrum}

We now consider a general magnetic field spectrum
with cutoff $k_d$ much
smaller than the Lyman-Werner or X-ray momentum scales. This includes
the case of power-law spectra produced by turbulent decay of a 
magnetic field in a 
plasma. The original field may have been generated
by the mechanisms of \cite{BFJ} or 
\cite{Nirmalya} or, for example, by phase transitions
in the early Universe \cite{Durrer}.

The power spectrum of the source term is given by the convolution
\begin{equation}
\label{eq:genA2}
    S^2(k) = \gcoup^2 \int d\bo{\Tilde{k}}^3
        B^2(\Tilde{k}) \delta \phi'^2(|\bo k - \bo{\Tilde k}|).
\end{equation}

\noindent Since $B^2(\Tilde{k})$ has a cutoff at $k_d$, the integral 
only has support in a region where $\Tilde k \leq k_d \ll k$. Therefore
$|\bo k - \bo{\Tilde k}| \simeq k$ up to a term of order 
$k_d \ll k$.\footnote{If $k_d \sim 1 \, \rm{pc}^{-1}$ and $k$ is in the
Lyman-Werner or X-ray range, one has $k_d/k \lesssim 10^{-24}$.}
Hence, $\delta \phi'^2(|\bo k - \bo{\Tilde k}|)
\simeq \delta \phi'^2(k)$ and we can reduce the convolution to the product\footnote{The 
correction is controlled by the parameter
$k_d (\eta-\eta_{rec})$. The extra
term in $S^2(k)$ grows at most to order 1 and has the same scaling 
with $k^{-1}$.
It therefore doesn't affect the approximation \eqref{large-limit2}.}
\begin{equation}
\begin{aligned}
   S^2(k) &\simeq \gcoup^2
        \delta \phi'^2(k) \int d\bo{\Tilde{k}}^3 B^2(\Tilde{k}) \\
        &\sim \gcoup^2 \delta\phi'^2(k) \rho_B,
\end{aligned}
\end{equation}
where $\rho_B$ is the total energy density of the
magnetic field. Making use of \eqref{phik}, we find for the
source term
\begin{equation}
\begin{aligned}
    S_i(k) &\sim \gcoup \delta\phi'(k) B_0 \\
        &= \gcoup \frac{1}{2}\Phi B_0 k^{-1/2} \mathcal{A}^{1/2}
       \cos k(\eta - \eta_{rec}),
\end{aligned}
\end{equation}
where $B_0 \equiv \sqrt{\rho_B}$. 
Substitution of this expression in \eqref{Aeq} gives again, for $k \gg k_d$:
\be \label{large-limit2}
A_i(k) \, \sim \, g_{\phi \gamma} \Phi {\cal{A}}^{1/2} B_0 \frac{1}{k^{5/2}}.
\ee

\section{Comparison with Observations}

In this section we ask whether it is possible to get a sufficient flux of Lyman-Werner photon (in order to open up the Direct Collapse Black Hole formation channel) while at the same time being consistent with observational constraints.  We here consider three conditions: the first is that the flux of secondary photons in the CMB range be lower than the corresponding flux of primary CMB photons.  As a second criterion we take a first look at induced spectral distortions and demand that they are below the observational limit, and as a third we demand that the secondary flux of X-ray photons does not exceed the observed flux. Note that in \cite{Ashu} it was already shown that the flux of secondary photons is consistent with the results of ARCADE \cite{Arcade} and EDGES \cite{Edges} when these observations are interpreted as upper limits on the flux at radio wavelengths.

A first criterion is that the energy density in secondary photons on CMB anisotropy scales be smaller than the energy density in the CMB photons on these scales, i.e.
\be
\rho_A(k) \, \ll \, \rho_{CMB}(k) \, .
\ee
Since on these scales $k \ll k_c$ it follows from (\ref{small-limit}) that
\ba
\rho_A(k) \, &\sim& \, k^2 k^3 |A_i(k)|^2 \\
&\sim& \, g_{\phi \gamma}^2 \Phi^2 B^2 {\cal{A}} \frac{k}{k_c} \, , \nonumber
\ea
while at recombination
\be
\rho_{CMB}(k) \, \sim \, \frac{\delta T}{T} T_{rec}^4 \, \sim \, {\cal{A}}^{1/2} T_{rec}^4 \, .
\ee
Hence the ratio is 
\be
\frac{\rho_A(k)}{\rho_{CMB}(k)} \, \sim \, {\tilde{g}}_{\phi \gamma}^2 m_{20}^{-2} 10^{-2} \left( \frac{B}{1\, {\rm Gauss}} \right)^2\frac{k}{k_c} {\cal{A}}^{1/2} \, \ll \, 1
\ee
for the reference value $B = 1 {\rm Gauss}$ (since $k \ll k_c$).

The production of secondary photons will induce spectral distortions. The magnitude of these distortions are limited by the FIRAS observations \cite{FIRAS} which yield the condition
\be \label{crit2}
\frac{\Delta \rho_A(k_0)}{\rho} \, \ll \, 10^{-5} \, ,
\ee
where $\Delta \rho_A(k_0)$ is the energy density of photons with frequency $k_0$ in the range of the peak of the CMB spectrum.  Note that since the photons are non-interacting, it is only photons with wavelengths comparable to $k_0$ which contribute to spectral distortions.

Since now $k_0 \gg k_c$ we use (\ref{large-limit}) to obtain
\ba \label{density}
\Delta \rho_A(k_0) \, &\sim& \, g_{\phi \gamma}^2 \Phi^2 B^2 {\cal{A}} \\
&\sim& \, {\tilde{g}}_{\phi \gamma}^2 m_{20}^{-2} \left( \frac{B}{1\, {\rm Gauss}} \right)^2 10^{-2} {\cal{A}} T_{rec}^4 \, , \nonumber
\ea
and the criterion (\ref{crit2}) becomes 
\be
{\tilde{g}}_{\phi \gamma}^2 m_{20}^{-2} \left( \frac{B}{1\, {\rm Gauss}} \right)^2 \, \ll \, 10^4 \, ,
\ee
where we have inserted the COBE-normalized value ${\cal{A}} \sim 10^{-9}$. 

We see that for the reference values of the parameters which we are using,  the induced spectral distortions will be below the observational limits. We realize that the above is a very qualitative analysis and should be followed up by a careful numerical study, e.g. using the code of \cite{Chluba}.

The third criterion which we consider here is the X-ray flux.  The secondary X-ray photons generated via our mechanism cannot produce a larger X-ray flux than observations \cite{Chandra} indicate. In the range of about $1 {\rm keV}$ the bound 
at recombination is
\begin{equation}
\begin{aligned}
    \label{bound}
    \Phi_A(k) \, \ll 10^{-49} {\rm GeV}^3 a_{rec}^{-3} \, 
    \sim 10^{-40} \, {\rm GeV}^3.
\end{aligned}
\end{equation}
The X-ray flux which our mechanism produces is
\be \label{prediction}
\Phi_A(k) \, \sim \, \frac{\rho_A(k)}{k} \, \sim \, {\tilde{g}}_{\phi \gamma}^2 m_{20}^{-2} \left( \frac{B}{1 \,{\rm Gauss}} \right)^2 10^{-41}  {\rm GeV}^3
\ee
where in the second step we have made use of (\ref{density}). The consistency between the bound (\ref{bound}) and the prediction (\ref{prediction}) requires parameter values which satisfy
\be
{\tilde{g}}_{\phi \gamma}^2 m_{20}^{-2} \left( \frac{B}{1 \,{\rm Gauss}} \right)^2 \, \ll \, 10 ,
\ee
which is not hard to justify since observations indicate that $m_{20} \gg 1$ and ${\tilde{g}}_{\phi \gamma} \ll 1$.
Hence, the observational upper bounds tested in this section 
are consistent with the Lyman-Werner flux required
to open the DCBH channel
\be \label{require}
\Phi_A(k) \, > \, 10^{-44} {\rm{GeV}}^3 \,, 
\ee
\noindent which demands
\be 
{\tilde{g}}_{\phi \gamma}^2 m_{20}^{-2} \left( \frac{B}{1 \,{\rm Gauss}} \right)^2 \, > \, 10^{-5}.
\ee

\section{Discussion}

We have considered the production of secondary photons from the coupling between a magnetic field $B$ present on cosmological scales at recombiination and the COBE-normalized dark matter fluctuations, assuming that the dark matter is a pseudo-scalar axion-like particle which couples to electromagnetism via the usual Chern-Simons coupling.  In a previous work \cite{Ashu} we had assumed that $B$ is homogeneous. Here, we relax this assumption and study the case where $B$ consists of a spectral distribution of cosmological scale modes.  We fiind that it is possible for our mechanism to produce a sufficient flux of Lyman-Werner photons (to allow for direct collapse black hole formation) while being consistent with the constraints on the flux of secondary photons from the CMB and from X-ray observations.

Our analysis is applicable independent of what the source of the cosmological magnetic field present at the time of recombination is, although our work was motivated by the mechanism of producing a cosmological scale magnetic field from a parametric resonance of the electromagnetic field after recombination in the presence of a coherently oscillating pseudo-scalar field with the same Lagrangian as we consider here \cite{BFJ, Nirmalya}.

\section*{Acknowledgement}

\noindent 
 
AA thanks the Department of Physics of McGill University for its hospitality. His work is supported in part by funds from CAPES - 001. RB thanks Nirmalya Brahma, J\"{u}rg Fr\"{o}hlich,  Hao Jiao and Vahid Kamali for discussions and for collaboration on related works. The research at McGill is supported in part by funds from NSERC, from McGill University and from the Canada Research Chair program.  The work of AK was supported by the Japan Society for the Promotion of Science (JSPS) as part of the JSPS Postdoctoral Program (Grant Number: 25KF0107).

\end{document}